\documentclass[electronic]{vgtc}             

\ifpdf
  \pdfoutput=1\relax                   
  \usepackage{graphicx}                
  \DeclareGraphicsExtensions{.pdf,.png,.jpg,.jpeg} 
\else
  \usepackage{graphicx}                
  \DeclareGraphicsExtensions{.eps}     
\fi%

\graphicspath{{figures/}{pictures/}{images/}{./}} 

\usepackage{microtype}                 
\PassOptionsToPackage{warn}{textcomp}  
\usepackage{textcomp}                  
\usepackage{mathptmx}                  
\usepackage{times}                     
\usepackage{cite}                      
\usepackage{tabu}                      
\usepackage{booktabs}                  
\usepackage{longtable}
\usepackage{float}
\usepackage[hyphens]{url}
\usepackage{graphicx}
\usepackage{subcaption}
\usepackage{color}
\usepackage[normalem]{ulem} 
\usepackage{etoolbox}
\newtoggle{showcomments}
\togglefalse{showcomments} 
\newtoggle{showedits}
\togglefalse{showedits} 
\newtoggle{showadditions}
\togglefalse{showadditions}

\newcommand{\added}[1]{\iftoggle{showadditions}{\textcolor{blue}{#1}}{\textcolor{black}{#1}}}
\newcommand{\deleted}[1]{\iftoggle{showedits}{\textcolor{red}{\sout{#1}}}{}}

\newcommand{\yuhang}[1]{\iftoggle{showcomments}{{\small\textcolor{red}{\bf [Yuhang: #1]}}}{}} 

\onlineid{1377}

\vgtccategory{Research}

\title{Springboard\added{, Roadblock} or \added{``}Crutch\added{''}?: How Transgender Users Leverage Voice Changers for Gender Presentation in Social Virtual Reality}

\author{Kassie C. Povinelli\thanks{e-mail: kassie.povinelli@wisc.edu}\\ %
        \scriptsize University of Wisconsin-Madison %
\and Yuhang Zhao\thanks{e-mail:yuhang.zhao@cs.wisc.edu}\\ %
     \scriptsize University of Wisconsin-Madison }%

\abstract{Social virtual reality (VR) serves as a vital platform for transgender individuals to explore their identities through avatars and foster personal connections within online communities. However, it presents a challenge: the disconnect between avatar embodiment and voice representation, often leading to misgendering and harassment. Prior research acknowledges this issue but overlooks the potential solution of voice changers. We interviewed 13 transgender and gender-nonconforming users of social VR platforms, focusing on their experiences with and without voice changers. We found that using a voice changer not only reduces voice-related harassment, but also allows them to experience gender euphoria through both hearing their modified voice and the reactions of others to their modified voice, motivating them to pursue voice training and medication to achieve desired voices. Furthermore, we identified the technical barriers to current voice changer technology and potential improvements to alleviate the problems that transgender and gender-nonconforming users face.
} 


\begin{document}


\firstsection{Introduction}

\maketitle

Social Virtual Reality (VR) platforms provide an immersive digital space for individuals to interact with each other, where a combination of technologies such as head-mounted displays, real-time voice chat,  full-body tracking, kinesthetic interactions, and traversable 3D worlds provide a Face-to-Face (FtF)-like interaction experience \cite{sense_of_presence_social_VR}. Increased adoption during the COVID-19 pandemic increases the importance of studying these interactions. In social VR, LGBTQ+ users find solace and empowerment through embodiment and community-building. They transcend physical limitations, forming connections with those who share similar experiences from across the world \cite{lgbtq_support_social_virtual_reality, safe_space_lgbtq_social_vr}.
While avatars in social VR serve as both representations of users and their interfaces for interacting with their environments and others, \textit{voice} is the primary channel that can expose a user' identity in real life \cite{rediscovering_the_body_online_trans, schulenberg2023creepy}. For many transgender and gender-nonconforming (TGNC) users whose voices may not align with their gender identity, even though they could customize avatars to present their preferred gender identity, the voice-avatar mismatch can make them targets of harassment within social VR spaces \cite{rediscovering_the_body_online_trans, acting_out_queer_identity}. Furthermore, \added{\textit{voice dysphoria}}---an emotional and psychological distress stemming from the incongruence between one's gender identity and voice---compounds the difficulties they face. 
\deleted{Voice training is a common method used by TGNC users to align their voice with gender identity. Voice training involves both active efforts to exercise vocal muscles and passive use of an altered but oftentimes nonpassing intermediary voice \cite{voice_training, oates2019evidence}. However, it is not always a viable solution since it can be a complex and arduous process, and some individuals may find it less preferred or even unattainable.}

\added{\textit{Voice training} is a common solution to issues of voice-based misgendering and voice dysphoria, but the considerable daily exercises and outward practice required can pose a formidable obstacle in overcoming voice dysphoria for numerous TGNC individuals striving to reconcile their voices with their gender identities. \yuhang{ANSWER: I'm not sure if this comment was from before I made the change, but this states that voice training is a solution --- this does not flow well by saying voice training is an obstacle directly... probably should start with voice training is an solution, introduce it, and then mention its issues.} Voice training involves both active efforts to exercise vocal muscles and passive use of an altered but oftentimes nonpassing intermediary voice in social communications \cite{voice_training, oates2019evidence}. While highly beneficial to TGNC individuals in the long-term, the overwhelming complexity, social risks of outwardly practicing (e.g., identity disclosure and harassment caused by speaking with intermediary voice), 
 and time-commitment required by voice training \yuhang{ANSWERED: (using freeman's work again here, shortly references some voice training material related to social VR and difficulties (such as worries that family will know if they are in an unaccepting household and they overhear)) need to reference for these drawbacks} prevents users from fully experiencing gender embodiment in social VR in the short-term \cite{rediscovering_the_body_online_trans}.}

As the challenges of voice representation become increasingly salient in social VR, the emergence of advanced voice changer technology offers a compelling solution. Voice changers hold particular significance for \added{TGNC} users who grapple with voice-avatar mismatches \added{\cite{rediscovering_the_body_online_trans}}\yuhang{ANSWERED: cite Freeman's papers}. The ability to modify their voices convincingly offers a pathway to better align their virtual identities with their preferred gender identities, reducing the risk of harassment and diminishing the emotional toll of voice dysphoria. Recent advancements in AI-based real-time voice generation have transformed the landscape  \added{\cite{Voicemod, MorphVOX_Pro, MorphVOX_Junior}}\yuhang{ANSWERED: reference}. As a result, the artificiality that once marred voice changers has been greatly reduced, allowing for more authentic and immersive experiences with voice in social VR.

Despite the increasing prevalence of voice changer usage, the transformative potential of these technologies in supporting TGNC users and addressing issues such as voice dysphoria and voice-based harassment remains largely unexplored. \added{There exists a notable research gap in understanding how voice changers are employed within the context of social VR, their impact on TGNC users' social experiences, and the technical and usability challenges faced by TGNC users when adopting these technologies.}

To address this research gap, we conducted a comprehensive interview study involving 13 participants from diverse backgrounds within the TGNC community in social VR. Our findings offer valuable insights into the ways in which these users have harnessed voice changers, shedding light on the multifaceted impact of this technology on their virtual and real-world experiences.
We found that the use of voice changers emerged as a transformative tool for many of our participants. It played a pivotal role in reducing instances of discrimination, bias, and hostility related to their identities in social VR environments. This newfound freedom allowed them to expand their boundaries and engage more comfortably in virtual social interactions.
Beyond mitigating external challenges, voice changers also proved instrumental in alleviating internal struggles. Participants reported a significant reduction in feelings of \added{voice} dysphoria. \deleted{a psychological distress stemming from the incongruence between one's gender identity and assigned sex.} The ability to hear a voice that resonated with their gender identity brought a profound sense of relief and congruence, bolstering their overall well-being.

While voice changers and voice training are the two typical methods used by TGNC users, our research revealed the tension and interaction between these two methods. Voice changers offered TGNC users the opportunity to ``pass'' in social VR, enabling them to be treated and perceived as their preferred gender identities. This experience, akin to a digital ``test run,'' provided a tangible demonstration of what passing could feel like, motivating them to explore voice training and other transition-related processes in the real world.
Even basic (non-AI) voice changers acted as valuable ``training wheels'' for voice training, offering users a stepping stone towards greater voice authenticity \added{through voice enhancements, rather than full conversions}. Additionally, the modified voice that users created oftentimes served as a goal for voice training. However, we also found that voice changers can become a roadblock to voice training\added{,} with \added{users highlighting several concerns. These include apprehensions about the perceptible change in their voice to others if they were to discontinue using voice changers, the feeling that AI voice changers create such drastic voice alterations that their own efforts seem less impactful, and worries about developing an overreliance on these tools as a ``crutch,'' which hinder consistent voice training.}\deleted{the users concerning their voice inconsistency to others} \yuhang{ANSWERED: inconsistency during the training process? need more clarification} 

Beyond the social impact, we also identified the \added{usability}\deleted{accessibility} issues of current voice changing technologies for TGNC users, regarding the technology complexity, capabilities, and affordability.

Our research provides \deleted{four}\added{five} contributions. First, this is the first research that studies the impact of voice changers on TGNC users' experiences in social VR. Second, we identify the goals of TGNC users when using voice changers and the factors influencing their decisions of when to use them. Third, we discuss the implications of passing as cisgender on gender exploration in social VR. Fourth, we derive design guidelines for effective and user-friendly voice changer applications. \added{Fifth, we develop the idea of voice changers in social VR as a potential ``springboard'' for voice training, framing it as the next step in incorporating technology into voice training and therapy regimens.}
\section{Related Works}
\yuhang{Related work is a long and thorough version to motivate our own research, see it as an extensive version of introduction. We are missing the piece that highlights the research gap and motivates our own research, across the whole related work section. See my comments in each subsection.}

\subsection{Marginalization of TGNC Users Online}
Marginalization refers to the systematic social exclusion, discrimination, and disempowerment of individuals or groups based on characteristics such as gender identity, sexual orientation, disability, race, or ethnicity \cite{marginalized_populations}. In the context of transgender and gender-nonconforming (TGNC) individuals, marginalization manifests in various forms, both online and offline, leading to detrimental consequences. The experiences of TGNC individuals in online spaces has been a subject of growing concern and investigation. Marginalization, discrimination, and harassment against TGNC users have been widely documented, shedding light on the challenges they face in digital environments \cite{social_media_trans_youth_wellbeing, harassment_social_media}. The marginalization of TGNC individuals both online and offline is associated with increased psychological distress, and increased coverage in media and legislation further increases feelings of stigmatization \cite{stigmatization_of_trans_people, media_messages_trans_stigma}. 

\paragraph{Community building} TGNC users, in response, create online communities for advocation, safety, and visibility. Haimson et al. explored how trans users created their own website, ``Trans Time,`` to share their trans experiences without fear of the stigmatization present on other social media platforms, while focusing on aspects such as safety, privacy, and content warnings \cite{trans_time_specific_social_media}. Jackson et al. explored how black transgender and queer women, using the ``\#GirlsLikeUs'' network on Twitter, engage in activism to counter narratives surrounding them online and in the media \cite{trans_women_twitter_community}.

\paragraph{Identity disclosure} TGNC users control the pieces of their identity that they display on social media. They control who they come out to, how much of their identity they disclose, and how they construct their online identities and streams to fit with their transgender identities \cite{gender_transition_facebook}. Haimson et al. explored how trans users use social media to come out on various forms of social media, such as blogs and Facebook pages, and the short and long-term impacts of disclosing identity on social media in comparison to real-life coming-out experiences \cite{trans_identity_disclosure_social_media}. Buss et al. investigated how transgender users engage in identity management on social media across platforms, finding that transgender users not only control their identity presentations through choosing what to post, but also curate their experiences by deciding what they see from others and who they interact with \cite{trans_identity_management}.

\subsection{Experiences of Vulnerable Users in Social VR}
Underrepresented users in social VR and virtual worlds frequently encounter discrimination, harassment, and exclusion based on their gender identity, sexual orientation, disability, race, or ethnicity. They may face offensive stereotypes, microaggressions, and outright hostility from other users, contributing to feelings of alienation and frustration \cite{ableist_migroagressions_disability, freeman2022disturbing}. Additionally, the lack of diverse representation in virtual environments can reinforce real-world disparities, perpetuating harmful biases and prejudices \added{\cite{impact_of_disability_signifiers, schulenberg2023creepy}}\yuhang{ANSWERED: need reference}. In response to these challenges, underrepresented users often form supportive communities, develop coping measures to counter stigmatization, and create new avenues for identity presentation within social VR and virtual worlds \added{\cite{safe_space_lgbtq_social_vr, lgbtq_support_social_virtual_reality, rediscovering_the_body_online_trans, disability_disclosure, avatar_self_representation_women_midlife}}\yuhang{ANSWERED: add representative references}.

\paragraph{Users with Disabilities}
Individuals with disabilities navigating social VR and virtual worlds often encounter accessibility issues that limit their full participation. These platforms may not provide adequate support for various disabilities, such as mobility impairments or sensory challenges. There is a body of HCI research focused on designing VR interfaces that meet the needs of people with disabilities (PWD) \cite{vision_impairment_vr_bubble_disability, vr_walking_stick_vision_impairment, disability_social_vr_design_accessibility}. \yuhang{ANSWERED (added transition sentence) -- need a transition sentence here. the following two works do not seem to be relevant to prior sentences} \added{Furthermore, users with disabilities oftentimes want ways to represent their disability identities in social VR.} Zhang et al. explored how PWD represent themselves in social VR using their avatars, both through how they disclose their disabilities and how they use avatars to communicate with others using techniques such as VR-ASL (a version of American Sign Language designed for the limited hand gestures of VR) \cite{disability_disclosure}. Further work identified the unique forms of verbal, physical, and environmental harassment that PWD face when disclosing their identities through the design of their avatars, including others mimicking their disabilities, insults, and others grabbing disability signifiers such as their wheelchairs and walking canes \cite{impact_of_disability_signifiers}.

\paragraph{\deleted{Female Users}\added{Women in Social VR}}
\deleted{Female users}\added{Women} in social VR and virtual worlds often face gender-based discrimination and harassment, reflecting broader societal issues. They may encounter unwelcome advances, objectification, or even verbal abuse from other users, and suffer from a lack of avatars representing their body types \cite{sexism_in_online_gaming, sexism_online_gaming_masculine_norms, avatar_self_representation_women_midlife}. These negative experiences can deter women from fully enjoying and participating in virtual communities and perpetuates gender disparities in these digital spaces. Schulenberg et al. found that social VR environments pose unique threats to women, since their embodied avatars in the 3D space become targets of physical harassment and the predominant use of voice communication makes hiding their identities difficult \cite{schulenberg2023creepy}.

\paragraph{LGBTQ+ Users}
LGBTQ+ users oftentimes lack sufficient offline social networks, and turn to online communities for information, support, and identity affirmation and formation \cite{lgbtq_youth_online_support, world_of_warcraft_identity_exploration, LGBTQ_visibility_social_media}. These online communities, however, lack aspects of FtF interactions, making them unsatisfactory when compared to offline support and community \cite{sexual_minority_women_online_dating, cmc_and_social_support}. Li et al. explored how the immersive and embodied experiences made possible through social VR empowers LGBTQ+ users by simulating offline support \cite{lgbtq_support_social_virtual_reality}. Acena and Freeman found that this embodied multi-modal interaction in social VR not only felt more supportive, but also translated to real life benefits \cite{safe_space_lgbtq_social_vr}.
\yuhang{need a sentence that connects to our research and contribution}
\subsection{Avatars for Identity Representation in Social VR}
\yuhang{ANSWER: it isn't quite clear how to merge the sections, since this section is specifically focused on identity presentation, which is only referenced in the PWD section of 2.2 -- this section feels repetitive to 2.2; should merge with 2.2, and adjust the title of 2.2 accordingly}
A burgeoning body of research examines users' utilization of avatars to depict themselves in social VR environments, as well as the consequences of avatar embodiment on their identities and interpersonal connections \cite{body_avatar_and_me_perception_of_self, long_distance_relationship_social_VR}. Avatars, through hand tracking and partial and full-body tracking, enable communication alternatives to voice, \added{ such as VR sign language \cite{disability_disclosure}, built-in avatar gesture recognition to communicate basic emotions and responses through avatar faces (a feature built into many VRChat avatars \cite{hai-vr_combo-gesture-expressions}), and using advanced avatar and world shaders such as virtual keyboards and chat boxes or virtual pens \cite{keyboard_girl_avatar_vrchat, vrchat_simple_pen_system}} \yuhang{ANSWERED: add references}. For marginalized users, such as transgender people and cisgender women, this enables them to communicate while retaining privacy around their voices, which can otherwise make them targets of harassment \cite{nonverbal_communication_social_virtual_reality}. In particular, transgender individuals utilize avatars within social VR platforms to synchronize their virtual representations with their preferred gender identities, providing a means for more intimate exploration and expression of their gender identities by bridging the gap to their physical bodies, while also empowering some to transition in real life \cite{rediscovering_the_body_online_trans}.
\added{
\subsection{Voice Training for TGNC Individuals}
Voice training plays a pivotal role in the lives of many TGNC individuals, serving as a key aspect of gender presentation and identity affirmation. 
}
\added{
It involves altering multiple vocal features, including pitch, resonance, intonation, and articulation. The process of voice training involves both physiological and pyschological adjustments, allowing TGNC individuals to increase their voice's congruence with their gender while also reducing the social stigma of having a gender-noncongruent voice, thus increasing quality of life \cite{ClevelandClinic2023TransVoice, voice_perceptions_QOL}. Physically altering one's voice may be achieved through individual exercises using online resources that compound videos, manuals, and regimens, such as r/transvoice's voice feminization and masculinization wikis, \cite{reddit_transvoice}, with the help of a speech therapist and regular visits combined with daily vocal exercises \cite{RCSLTTransVoice, SchneiderCourey2016}, or, in more complex cases, achieved through surgery on the vocal tract combined with intensive exercises afterward \cite{voice_feminization_surgery, SchneiderCourey2016}.
\paragraph{Voice Training for Transfeminine Individuals}
Voice training for transfemninine individuals often encompasses more than just pitch elevation; it involves a comprehensive approach addressing physical, neurophysiological, and acoustical aspects of voice \cite{patient_perspectives}. Morsomme and Remacle \cite{ambulatory_feedback} explored the impact of ambulatory biofeedback in helping a transgender woman increase her voice pitch in daily activities, suggesting its potential to extend new vocal behaviors beyond clinical settings. Gelfer and Tice \cite{voice_therapy_outcomes_15_months} conducted a study on the perceptual and acoustic outcomes of voice therapy for male-to-female transgender individuals, noting significant changes in how listeners perceived gender based on voice signals alone.
\paragraph{Voice Training for Transmasculine Individuals}
While voice feminization training is heavily covered academically, there isn't as much information on voice training for transmasculine individuals. This is because while feminizing hormone therapy for transfeminine individuals does not drastically change vocal characteristics, masculinizing hormone therapy for transgender men does \cite{SchneiderCourey2016}. Still, masculinizing hormone therapy on its own may not achieve some of the desired vocal results for transmasculine individuals, showing a need for vocal training alongside hormone therapy \cite{transmasc_hormone_therapy_voice}. Furthermore, some transmasculine individuals may not have access to or may not wish to use masculinizing hormone therapy. In these cases, voice training or therapy are the only methods available to them, since surgical techniques focus exclusively on feminizing voice \cite{SchneiderCourey2016}.
}\added{
\paragraph{Modern Voice Training Tools and Techniques}
The evolution of voice training has seen the development of mobile applications and online tools, providing accessible resources for transgender individuals seeking voice feminization or masculinization. Ahmed et al. \cite{voice_training_app} critically analyzed mobile voice training apps, highlighting their tendency to reinforce binary gender norms and societal expectations. The community-based design of free and open-source software for transgender voice training, as described by Ahmed et al. \cite{open-source_voice_training_app}, emphasizes the importance of user agency and collaborative development in creating effective voice training tools. Furthermore, Ahmed's study on trans competent interaction design \cite{trans_competent_interaction_design} sheds light on the need for inclusive and adaptive technologies that cater to the diverse goals and experiences of the transgender community.} 

\added{In addition to the vital but complex voice training, voice changers and social VR can potentially provide an easier and safer option for TGNC users to experiment and achieve their desired voice. Our research thus investigates TGNC users' adoption of voice changers, as well as the interaction between voice training and voice changers in social VR environments. }



\yuhang{I think we need a section for voice changers, where you introduce the current voice changing technology and start-of-the-art tools and APIs --ignore if we don't have space}

\section{Method}
To understand how TGNC users use voice changer technology to curate their identity in social VR, we conducted an in-depth interview study with \added{13} TGNC users who used voice changers in social VR.
\subsection{Participants}
We recruited \added{13} participants who are transgender or \deleted{gender non-conforming}\added{gender-nonconforming}, with ages ranging from 18 to 53 \deleted{($mean = 32.07$, $SD = 9.17$)}\added{($mean = 31.069$, $SD = 9.79$)}. \added{Participants were primarily located in the United States, with two participants located in Brazil and one in Germany.}
Our participants cover diverse gender identities, including ten transgender female \deleted{, one transgender male}, \deleted{one}\added{two} transmasc nonbinary and \deleted{two}\added{one} agender. All participants identified as transgender and five identified as gender-nonconforming. \added{Table \ref{tab:participants_demographics_1} shows participants' demographic information.} Furthermore, five participants reported having disabilities, including four \deleted{(P5, P9, P10, P12)}\added{(P4, P8, P9, P11)} identifying as neurodivergent (autism, ADHD) and one (\deleted{P6}\added{P5}) identifying as hard-of-hearing. One participant (\deleted{P12}\added{P11}) mentioned a stuttering disability that influenced her ability to speak with the voice changer. While using various types of voice changers (\added{Table \ref{tab:participant_experiences}}), all participants used VRChat as their primary social VR platform.

We designed our recruitment efforts to reach a diverse and representative sample of TGNC individuals who have engaged with social VR platforms and have experience using voice changers. We recruited participants through multiple channels, including Reddit (specifically the r/VRChat subreddit),  Discord (VRChat and VRC Trans Academy Discord servers), and various LGBTQ+ organizations that facilitated online postings. To be eligible for participation, individuals had to be at least 18 years of age, identify as transgender or gender-nonconforming, possess past or present experience with social VR platforms, have experience using voice changers within Social VR environments, and be fluent in English. \added{Before the formal study, we conducted a pilot study with a TGNC participant who had extensive technical knowledge and background in social VR development and voice changers. However, they did not use voice changer in social VR. We thus only used their data to refine the interview protocols and did not involve them in the final findings.}

\begin{table}[h!]
    \footnotesize
    \centering
    \caption{Participant demographics\added{, including Age, Race, Location (United States (US), Brazil (BR), or Germany (DE)), Gender (Female (F), Male (M), Agender (A), Nonbinary (NB)), whether the participant identifies as gender-nonconforming (GNC), and transition direction (Transfeminine (TF) or Transmasculine (TM))}\yuhang{ANSWERED: inconsistency in capitalization; the table also needs to be marked with color}}
    \vspace{-2ex}
    \begin{tabular}{p{.5cm}p{0.5cm}p{1cm}p{.7cm}p{1cm}p{.7cm}p{.7cm}}
        \toprule
        \textbf{ID} & \textbf{Age} & \textbf{Race} & \added{\textbf{Loc.}} & \textbf{Gender} & \added{\textbf{GNC}} & \added{\textbf{Trans. Dir.}} \\
        \midrule
        P1 & 53 & Caucasian & \added{US} & F & \added{No} & \added{TF} \\
        P2 & 28 & Caucasian & \added{US} & A & \added{Yes} & \added{N/A} \\
        P3 & 32 & Hispanic & \added{BR} & F & \added{No} & \added{TF} \\
        P4 & 20 & Caucasian & \added{US} & NB & \added{Yes} & \added{TM} \\
        P5 & 25 & Caucasian & \added{DE} & F & \added{No} & \added{TF} \\
        P6 & 22 & Caucasian & \added{US} & F & \added{No} & \added{TF} \\
        P7 & 18 & Brazilian & \added{BR} & F & \added{Yes} & \added{TF} \\
        P8 & 36 & Caucasian & \added{US} & F & \added{No} & \added{TF} \\
        P9 & 41 & Caucasian & \added{US} & F & \added{Yes} & \added{TF} \\
        P10 & 30 & Caucasian & \added{US} & F & \added{No} & \added{TF} \\
        P11 & 42 & Caucasian & \added{US} & F & \added{No} & \added{TF} \\
        P12 & 31 & Mixed & \added{US} & NB & \added{Yes} & \added{TM} \\
        P13 & 34 & Caucasian & \added{US} & F & \added{No} & \added{TF} \\
        \bottomrule
    \end{tabular}
    \label{tab:participants_demographics_1}
\end{table}

\subsection{Interviews}
We conducted a single-session interview study that lasted for one to two hours on Zoom. The study included four sections, focusing on participants' demographic information, social VR experience, voice chat experience, and technical issues with voice changers, respectively. We detail all interview questions in Appendix \ref{sec:appendix}. 

The first section focused on participants' demographic information. First, we asked about basic demographic information, including their age, ethnicity/racial background, gender identities (transgender identity, gender-nonconforming identity), and an open identity question. We then asked them about their real-life experiences with their voice, what voice changer they use or previously used in social VR, and how often they used the voice changer. Additionally, we asked participants if they encountered any social barriers in their real life because of their gender identity. For participants who had experiences with voice changers in social VR in the past, but no longer used them, we modified these questions to represent both their past and present experiences.

In the second section, we asked questions about participants' virtual world experiences, focusing on their experiences in social VR. We asked questions about their general social VR experiences, including what they usually did in social VR and the demographics of the people they hung out with. We asked participants how they presented themselves in social VR and for screenshots of their avatars. We asked various questions about participant's avatars, including questions on gender presentation and effectiveness of gender presentation if they indicated that the avatar represented their gender, and if they noticed differences in how other users treated them depending on the avatar they used if they indicated using multiple differing avatars. Participants also discussed both the benefits and barriers of using social VR as a TGNC user (depending on their identity).

The third section focused on using voice chat in social VR, and was split into two subsections. In the first subsection, we asked participants about experiences when using their \textit{unmodified voice} (that is, without a voice changer), including who they used their unmodified voice with, why they used their unmodified voice in certain scenarios, and the impacts of using unmodified voice. For participants who avoided using unmodified voice, we asked them about the strategies they used to avoid communicating with their unmodified voice. In the second subsection, we asked participants about their voice changer use, including how they discovered and started using voice changers, how voice changers influenced their social VR experiences, and who they decided to turn the voice changer off with (if at any time they did). We also asked participants to send samples of their \textit{modified voice} (that is, with a voice changer) or use their modified voice in the Zoom interview, and discussed the properties of the modified voice and how they tuned the voice changer to achieve a desired voice.

In the final section, we asked questions covering any technical aspects of the voice changer that were not mentioned in the previous section, including how the voice changer worked and the benefits and limitations of the voice changer software. We also asked how they wanted to improve the voice changer software. We finished the interview with a brainstorming session, where the interviewee and participant discussed possible social VR-based interfaces for voice changers, including implementation into the menu and avatar selection interfaces. All participants were compensated at a rate of \$25/hour upon the completion of the study. 
\subsection{Data Recording and Analysis}
Upon obtaining informed consent from all participants, we recorded all \added{13 participants'} interviews using Zoom's recording feature. These recorded interviews were then automatically transcribed by Zoom's transcription service. To ensure the accuracy and fidelity of the transcriptions, one researcher manually reviewed the transcripts while listening to the corresponding audio. Participants were given the flexibility to use or turn off their cameras during the interviews.

\deleted{For the analysis}\added{When analyzing participant transcripts}, we employed thematic analysis \cite{thematic_analysis}, a standard qualitative research method that aids in identifying recurring patterns and themes within the collected data. \added{As opposed to quantitative statistical analysis, this method enables us to ground our analysis directly in participants' complex experience, and capture concepts, opinions, and the rationales behind them via bottom-up approaches, by highlighting and summarizing key information from the rich textual data with low-level codes and categorizing the codes into high-level themes to generate recurring patterns or findings.}

Specifically, we selected two representative samples from the set of interviews. Two researchers coded the samples independently using open-coding and discussed and reconciled their codes, developing an initial codebook. Next, one researcher, using the codebook, coded the rest of the transcripts, and added new codes as needed \added{upon the agreement between two researchers}. 
\added{We then categorized all the codes into high-level themes and subthemes using axial coding and affinity diagram. After the initial themes were identified, researchers cross-referenced the original data, the codebook, and the themes, to make final adjustments, ensuring that all codes fell in the correct themes.}
The \deleted{final}\added{initial} codebook contained over 600 codes, which the researchers categorized into 11 themes. \added{We then removed non-novel themes that had already been explored in previous works, and included 6 prominent and novel themes with 23 subthemes. Tables \ref{tab:theme} and \ref{tab:theme2} in Appendix shows all themes, subthemes, and codes from our analysis. We elaborate all themes as our findings in Section \ref{sec:findings}.}

\subsection{Positionality Statement}
We recognize that our personal identities and cultural backgrounds inevitably intersect with our roles as researchers and may influence various aspects of our work. It is crucial to acknowledge these intersections to ensure transparency and a comprehensive understanding of the study's context and potential biases.

The primary researcher for this study is a transgender woman. Her lived experience as a member of the transgender community played a pivotal role in shaping the research. As the sole interviewer, she brought her firsthand understanding of transgender and gender-nonconforming experiences to the forefront. This informed her approach to interview design, the development of questioning routes, and the nuanced interpretation of interview data.

The second researcher on our team is a cisgender woman with a background in social VR research. Her expertise in the field contributed to maintaining the quality and rigor of our research materials and procedures. She provided valuable insights into the broader landscape of social VR, ensuring that our study aligns with existing research and effectively addresses pertinent issues.

\section{Findings} \label{sec:findings}

\subsection{Voice Changer Use Demographics}
Our participants utilized various voice changer software options, including \textit{Voicemod} \cite{Voicemod}, a tool that offers both AI and basic voice-changing capabilities, \textit{Clownfish} \cite{Clownfish}, a more basic voice-changer that directly alters the audio on a frequency-level, \textit{MorphVOX} \cite{MorphVOX_Pro, MorphVOX_Junior}, a voice changer that uses neural networks for voice modification, \textit{SuperVoiceChanger} \cite{SuperVoiceChanger}, a basic, free piece of voice changing software, and \textit{W-okada} \cite{w-okada_voice_changer}, an open-source AI-based voice changer. Notably, Voicemod proved popular \added{with six participants using it}: \deleted{P5 and P12}\added{P4 and P11} chose Voicemod Pro, \deleted{P2, P3, and P11}\added{P1, P2, and P10} opted for the non-Pro version, and \deleted{P10}\added{P9} favored the Voicemod Persona AI voice feature specifically. Clownfish emerged as another common choice, selected by five participants \deleted{(P3, P4, P6, P7, P9)}\added{(P2, P3, P5, P6, P8)}. \deleted{P13}\added{P12} used MorphVox Pro, while two participants \deleted{(P9, P14)}\added{(P8, P13)} used MorphVox Junior, and \deleted{P8}\added{P7} selected SuperVoiceChanger. This diversity highlights the wide array of voice changer tools participants actively employed.

\begin{table}[h!]
    \footnotesize
    \centering
    \caption{Participant Experiences with Social VR and Voice Changers \yuhang{ANSWERED: what does the star mean for p11? need explanation in caption. The modification to the table needs to be marked with color}}
    \vspace{-5ex}
\begin{center}
\begin{tabular}{p{.3cm}p{1.4cm}p{2.6cm}p{2.5cm}}
\toprule
\textbf{ID} & \textbf{Voice Changer Experience} & \textbf{SocVR Frequency} & \textbf{Voice Changer} \\
\midrule
P1 & Current & 4-7 hours, 4 days/week & Voicemod \\
P2 & Current & 3-6 hours, 5 days/week & Voicemod, Clownfish \\
P3 & Past & \added{3-8 hours daily} & Clownfish \\
P4 & Current & evenings, \newline 2-3 days/week & Voicemod Pro \\
P5 & Current & weekly & Clownfish \\
P6 & Current & 6-12 hours/week & Clownfish \\
P7 & Current & 2-3 hours daily & SuperVoiceChanger \\
P8 & Past & 1-3 hours daily & Clownfish, MorphVOX Junior \\
P9 & Current & variable & Voicemod Persona \\
P10 & Current & \added{weekly} & Voicemod \\
P11 & Current & limited & Voicemod Pro \\
P12 & Past & 2-8 hours/week & MorphVOX Pro \\
P13 & Past & 4-9 hours/week & MorphVOX Junior \\
\bottomrule
\end{tabular}
\end{center}
\label{tab:participant_experiences}
\vspace{-5ex}
\end{table}

In terms of the frequency of voice changer use, a notable trend emerged, with the majority of participants indicating near-constant utilization. Specifically, eleven participants (all except \deleted{P1, }\deleted{P3}\added{P2}\deleted{,} and \deleted{P11}\added{P10} \yuhang{P2 and P10 seems to use it frequently based on Table 2; ANSWER: Table 2 tracks frequency of social VR use, but not voice changer use}) reported using voice changers almost all the time, underlining the integral role of voice changers in their social VR interactions. Furthermore, participants demonstrated \added{differences} \deleted{discerning choices} in terms of whom they \deleted{are}\added{were} willing to use their unmodified voices with in social VR. \deleted{While many participants consistently employed voice changers,}\deleted{four}\added{Eight} participants \deleted{(P5, P7, P9, P12)}\added{(P2, P3, P4, P5, P6, P7, P8, P11)} expressed a willingness to use their unmodified voices with some \added{or all of their} close friends. On the other hand, \deleted{three}\added{five} participants \deleted{(P10, P13, P14)}\added{(P1, P9, P10, P12, P13)} \deleted{used a}\added{never turned off their} voice changer even \added{when} with close friends. These findings highlight the complex interplay between voice changer usage, privacy, and interpersonal dynamics within the social VR landscape.

\added{
Interestingly, P10 used voice changers all the time in non-VR virtual social worlds, including roleplay environments like \textit{Grand Theft Auto Online} \cite{gta_online} role-play servers and \textit{Conan Exiles} \cite{conan_exiles}\yuhang{ANSWERED: add reference for these two games}, and other voice chat reliant games, but used voice changers minimally in social VR, estimating use at only 10\% of the time. She regarded VRChat, her main social VR application, as a space for voice training, and thus tried to minimize her use of voice changers in the environment. 
}
\subsection{Motivations of Using Modified Voice} \label{sec:findings:motivations}
Participants approached the utilization and customization of voice changers with a range of distinct objectives in mind. While a common thread among all participants was the desire to authentically express their gender identity, many articulated specific motivations for harnessing this technology, including (1) adopting voice changers as a less arduous alternative to traditional voice training for achieving gender-affirming vocal presentations, (2) employing voice changers as a protective measure against potential instances of voice-related harassment, (3) aligning their virtual voice with their avatar's gender presentation, and (4) achieving an increased sense of personal comfort and confidence during social interactions in social VR. We elaborate on these motivations below.
\paragraph{Less Effort for Gender Identity Presentation} \label{sec:findings:less_effort}
Six participants \deleted{(P4, P6, P7, P10, P13, P14)}\added{(P3, P5, P6, P9, P12, P13)} acknowledged that voice changers offered them a more manageable alternative to the demanding processes of medical transition and rigorous voice training to achieve their desired vocal presentation of their gender identities.\added{ While having limited support in current social VR platforms, and thus not a full solution to voice difficulties, voice changers offered our participants an accessible voice matching their identities.} For transgender women and transfeminine nonbinary individuals \added{(all participants other than P2, P4, and P12)}, feminizing hormone therapy (MtF HRT) does not change their voice, and achieving a feminine voice through voice training can be extremely difficult, as training requires constant vocal exercises and safe spaces where they can practice using their trained voices \yuhang{ANSWERED: participant number?}. For the transmasculine participants \deleted{(P5, P13)}\added{(P4, P12)}, Testosterone-based HRT provided a route to an affirming voice, but was difficult to acquire as a result of medical gatekeeping and financial costs. When \deleted{P5}\added{P4}'s voice started dropping because of testosterone, he no longer felt a need for the voice changer.

\paragraph{Avoiding Harassment}
Six participants \deleted{(P5, P6, P9, P10, P12, P13)}\added{(P4, P5, P8, P9, P11, P12)} openly discussed the harassment they endured due to their unmodified voices in social VR. Three participants \deleted{(P3, P5, P9)}\added{(P2, P4, P8)} explicitly mentioned that even if they used an avatar aligning to their gender identity, others would often determine their gender based on their voice. The adoption of a voice changer offered an avenue to evade both the stigmatization directed at transgender users in social VR and, in the case of participants \deleted{transmasculine}\added{transitioning to a more masculine identity} \deleted{(P5, P13)}\added{(P4, 12)}, the harassment typically directed at \deleted{females}\added{women}.

\paragraph{Matching Voice to Avatar}
\deleted{Nine}\added{Eleven} participants \deleted{(P4, P5, P6, P7, P8, P9, P10, \added{P11,} P13, P14)}\added{(except for P1 and P11)} mentioned avatar-voice mismatch as a concern when presenting their gender identities and interacting with others in social VR. Using a voice changer to match their voice with the avatar reduced harassment from others, alleviated dysphoria, and made them more comfortable in their online identities. \deleted{P9}\added{P8}, who also uses her avatar for streaming from VRChat, mentioned how her unmodified voice doesn't sound right coming from her avatar, leading to \added{voice-based} gender dysphoria:
\added{``}\textit{I really do feel like my voice, my actual voice does not match my avatar. It's really dysphoric. When I think about it, I would much rather have a much more feminine voice. And it really, it really makes it hard to look at my VODs when I'm streaming because I'm hearing my voice coming from an anime girl and it's just like, yeah, why can't you just be more feminine?}\added{''}
\paragraph{Confidence in Voice}
Three participants \deleted{(P4, P8, P10)}\added{(P3, P7, P9)} mentioned that using a voice changer increased their confidence when speaking with others, while five \deleted{(P5, P7, P9, P10, P12)}\added{(P4, P6, P8, P9, P11)}\yuhang{ANSWERED: fix the number?} participants stated that they would not interact with strangers in social VR without a voice changer. \deleted{P10}\added{P9} highlights this increase in confidence:
    \added{``}\textit{[T]hey're not hearing me, they're hearing a more idealized voice they're hearing, it will change how I feel about myself. Because it's not me they're hearing. And in my eyes, that's fine. And that's what enables me to engage like I do, because they're not hearing me.}\added{''}
\subsection{Social Affirmation of Identity via Passing} \label{sec:findings:social_affirmation}
Ten participants \deleted{(except for \deleted{P1, }P3, P7, P11)}\added{(except for P2, P6, P10)} with binary-related identities (that is, they identified with or in proximity to a binary identity) mentioned that voice changers enabled them to have experiences with \textit{passing}---the social feedback of being treated as cisgender in their gender identity. These experiences of passing differ from just being around individuals that gendered them appropriately, in that even the unconscious biases separating transgender and cisgender identities disappeared.

Social feedback from others can play a significant role in affirming the gender identities of transgender individuals. Utilizing a passing voice changer cultivates an environment in which transgender users consistently receive affirmation of their gender identity, as others perceive and treat them in alignment with that identity. For several participants \deleted{(P4, P6, P8, P9, P12, P14)}\added{(P3, P5, P7, P8, P11, P13)}, the experiences of being perceived as cisgender and hearing themselves with a passing voice served to reinforce their gender identities. Additionally, for three participants \deleted{(P2, P13, P14)}\added{(P1, P12, P13)}, the social feedback they received from the use of voice changers contributed to their process of self-discovery and bolstered their confidence in their new gender identities.
\subsection{Experiences of Sexism and Preferential Treatment} \label{sec:findings:sexism_preferential_treatment}
Four participants \deleted{(P5, P8, P13, P14)}\added{(P4, P7, P12, P13)} mentioned experiences with sexism and preferential gender treatment in social VR as a result of passing as cisgender. 
For the transmasculine participants \deleted{(P5, P13)}\added{(P4, P12)}, using the voice changer swapped their experiences from preferential treatment and sexism to being treated as ``one of the guys.'' For transfeminine participants \deleted{(P8, P14)}\added{(P7, P13)}, using the voice changer created situations where they were targets of sexism or preferentially treated by male VRChat users.

\paragraph{Transmasculine Experiences}
\deleted{P5 and P13}\added{P4 and P12} mentioned that when not using the voice changer, they became targets of sexism and harassment based on their perceived genders. Other users perceived them as \deleted{female}\added{women}, even when they were using male avatars. \deleted{P13}\added{P12} described experiences of harassment when using their unmodified voice in social VR: 
 \added{``}\textit{I was immediately like, gendered in that phrase, it was like someone goes, someone insulted... They were like, they referred to me as bitch.}\added{''}

\deleted{P5}\added{P4} highlighted that using a voice changer makes sexist harassment far less frequent, allowing him to be taken seriously by male users:
    \added{``}\textit{I no longer experience instances like that kid coming up to me and asking me why I was a woman in that [male] avatar. People make less sort of comments about me that are in that nature. If someone's gonna make fun of me, it's because I did something ridiculous in VR as a joke. And not because of how I sound or my identity.}\added{''}

Both \deleted{P5 and P13}\added{P4 and P12} had experiences of inclusion in male-dominated social circles, where utilizing voice changers allowed them to navigate and interact within these spaces. \deleted{P5}\added{P4} articulated this phenomenon, stating:
    \added{``}\textit{[P]eople are more comfortable with including me or like having myself sort of insert my own take to the joke more like, I find that the public worlds guys will primarily want to hang out with other guys and will be a bit more rude to women and so when I'm using the voice mod I get included in that guys group.}\added{''}
    
Both \deleted{P5 and P13}\added{P4 and P12} recognized a loss in preferential treatment, with \deleted{P13}\added{P12} experiencing harassment from males because his modified voice, while passing as male, sounded younger than his age. Both participants viewed this loss in preferential treatment positively. As \deleted{P13}\added{P12} highlighted, preferential treatment based on being perceived as \deleted{female}\added{a woman} is a source of disgust:
    \added{``}\textit{I actually feel disgusted when I feel like I'm getting special treatment from cis guys, just because they perceive me to be a female or feminine.}\added{''}
\paragraph{Transfeminine Experiences}
For transfeminine participants, using voice changers often led to the barriers of being perceived as cisgender women in social VR environments. While this enabled them to explore a gender presentation aligned with their gender identities, it also exposed them to sexist behavior and harassment. \deleted{P14}\added{P13} shared insights into how presenting as cisgender \deleted{female}\added{woman} with a voice changer brought about both positive and negative experiences:
    \added{``}\textit{It will also reveal to you what the negatives of being your preferred gender are very quickly. You know, people think you're a girl and they find you attractive. Guys of all stripes will try to interact with you very poorly, and in a wide range of ways.}\added{''}
\subsection{Voice Changers and Voice Training} \label{sec:findings:changers_and_training}
\yuhang{ANSWERED: we should reframe this section to explain how voice changer could be a springboard, a roadblock, and a crutch... nonw the roadblock and crutch are mixed together, but they mean differently. }
Participants held divergent perspectives regarding the relationship between voice changers and voice training. Some participants perceived voice changers as impediments to their transition, viewing them as a \added{roadblock. Others thought voice changers offered useful support, but could cause overreliance, acting as a} ``crutch.'' \yuhang{ANSWERED: do you mean roadblock? need to distinguish these two conditions} Conversely, many participants regarded voice changers as valuable aids in their voice training journey. For certain individuals, the altered voice served as a benchmark for their voice training goals, while for others, particularly those using simpler, non-AI voice-changing technologies, the voice changer acted as a form of vocal ``training wheels.''
\paragraph{Voice Changers as a \added{Roadblock}}
\deleted{P4, P10,\added{ and P11}}\added{P3, P9, and P10} felt that voice changers posed certain roadblocks to voice training. \deleted{P10}\added{P9} split voice training into two categories: passive voice training, where a transgender individual practices their voice by using it regularly in conversation, and active voice training, which includes vocal exercises and listening to a recording of one's voice through headphones. The use of voice changer can largely discourage a transgender user's passive voice training. \added{P9 stressed that as the ``almost ideal'' voice} from \added{the} voice changer becomes one's identity in social VR, and passive voice training necessitates the use of the \added{partially trained but intermediary voice}, overcoming the fear of others hearing a different, \added{``non-ideal''} voice \added{in passive voice training} became a significant challenge for those reliant on voice changers:
     \added{``}\textit{[P]eople have gotten used to your real voice. And they're now hearing your altered voice like your intentionally changed sound. So it's a lot like dropping the Voicemod and using a different voice.}\added{''}
     
\paragraph{\added{Voice Changers as a ``Crutch''}}
\added{\deleted{P11}\added{P10}, while using voice changers heavily in non-VR social online environments, made an effort to minimize her usage of voice changers in social VR. Treating social VR as an opportunity to practice using her new voice, P10 felt that voice changers sometimes acted as a barrier, but other times provided a necessary crutch---a support that can cause overreliance:} 
    \added{``\textit{The better term [to describe the voice changer] might be a crutch. Because I can always run back to this type of scenario... It lowers my motivation. If I keep it [on] it would lower my motivation if I rely on it too much. In certain scenarios, if that makes sense. [It lowers] my motivation to learn and, and work on my voice. The fact that it's there is always good for if I'm too tired to even try.}''}

\added{P5 highligheted how, while useful, her overreliance on an AI voice changer prevented her from voice training in social VR. While the voice changer allowed her to access an affirming voice in social VR, she acknowledged that voice trainining would be necessary to affirm her identity outside the virtual world.}
    \added{``\textit{I've known I would have to do [voice training] since the very beginning, but I just didn't because I thought you know, have my computer adjusted for me. Which you know, as you get older, life exists outside of computer screen, and suddenly you can't do that anymore.}''}  
\added{However, P5 also mentioned that this issue did not pertain to basic voice changing software like Clownfish, which could only change pitch and thus required voice training efforts alongside them.}

\paragraph{Voice Changers for Setting Voice Training Goals}
Five participants \deleted{(P6, P8, P10, P12, P14)}\added{(P5, P7, P9, P11, P13)} regarded their modified voices by the voice changer as either a voice training goal or an ideal vocal representation they aimed to achieve. By establishing a target voice, particularly in situations where the voice changer retained elements of their natural voices, it served as a tangible model for attainable vocal results through training. \deleted{P6}\added{P5}, for instance, expressed:
\added{``}\textit{[The modified voice] almost gave me a goal to strive for. And it gives me the feeling of what it's like to actually pass on the internet. And that is invaluable.}\added{''}


\paragraph{Voice Changer as ``Training Wheels'' for Voice Training}
Six participants \deleted{(P2, P4, P6, P7, P9, P14)}\added{(P1, P3, P5, P6, P8, P13)} trained their original voice alongside the voice changer, with social VR as a training ground. Of these participants, all were using voice changers that preserved aspects of their original voice, rather than synthesizing an entirely new one. \deleted{P2}\added{P1} employed a basic voice changer to solely adjust the pitch of her voice. Gradually, she reduced the pitch modification while simultaneously refining her voice through training. Over time, she succeeded in lowering the pitch modification to a point where she felt comfortable turning off the voice changer entirely, relying solely on her trained voice. This approach circumvented the challenges associated with using an unmodified voice, while also mitigating the abrupt and noticeable changes in one's vocal presentation to others mentioned by \deleted{P10}\added{P9}. \added{For P5, the limitations of non-AI voice changers like Clownfish, which could only change pitch, opened up new avenues for voice training, since training other aspects of voice, such as resonance, could be done alongside the voice changer, which would offer a necessary support when focusing on specific areas of voice training:
    \added{``}\textit{[Y]ou have to practice it whilst listening to how it's changing your voice in real time. It's almost like enhancement, rather than a converter, if that makes sense.}\added{''}
}

\subsection{Limitations of Voice Changers} \label{sec:findings:limitations_of_voice_changers}
Several participants encountered various technical challenges while configuring and using their voice changers. These difficulties encompassed complex and convoluted settings that impeded a clear understanding of the voice changer's functionality. Many participants also grappled with voice changer artifacts that not only caused annoyance to others but also risked exposing their true identities. Additionally,\deleted{ machine} intelligibility issues\added{, which occured when others could not understand them when the voice changer was in effect, either because of glitchiness, low bitrate, or other voice artifacts,} further complicated their experiences. Those who experimented with AI-based voice changers reported performance problems stemming from system strain, further exacerbating the technical hurdles they faced.

\paragraph{Difficulties with Settings}
Several participants encountered significant challenges during the configuration of their voice-changing software. Specifically, three participants \deleted{(P7, P9, P12)}\added{(P6, P8, P11)} expressed frustration with the complexity of voice changer settings. \added{These interfaces were often made of multiple interacting components controlled by sliders, with little connection to real-life vocal characteristics like resonance or weight, as shown in the interfaces of SuperVoiceChanger (Figure \ref{fig:supervoicechanger}), VoiceMod (Figure \ref{fig:voicemod}), and Clownfish (Figure \ref{fig:clownfish}).} \deleted{P4 and P9}\added{P3 and P8} detailed their experiences of discovering functional voice changer settings through trial-and-error, while \deleted{P4 and P5}\added{P3 and P4} noted their reliance on assistance from friends in this process. Moreover, three participants \deleted{(P4, P8, P13)}\added{(P3, P7, P12)} voiced concerns regarding the available voice presets. \added{Examples of these pre-loaded presets include MorphVox's ``Man'', ``Woman'', and ``Tiny Folks'' voice presets (Figure \ref{fig:morphvox}), and the pre-loaded anime-style presets of w-okada (Figure \ref{fig:wokada}).}
For example, \deleted{P13}\added{P12} highlighted the lack of humanlike voice changers in his search for presets,
    \added{``}\textit{[A] lot of options just sounded like caricatures of a voice or something like that. Like an animated or like inhuman comedic or some, like very animated version of an actual human voice.}\added{''}

\paragraph{AI Voice Changers}
Four participants \deleted{(P4, P6, P10, P12)}\added{(P3, P5, P9, P11)} raised concerns regarding the compatibility of AI-based voice changers, such as Voicemod Persona, with VRChat. They reported instances where both VRChat and the AI voice changer struggled to efficiently allocate system resources, resulting in undesirable consequences such as audio glitches and robotic-sounding voices. These performance disruptions manifested as various issues, ranging from users detecting the use of a voice changer to complete voice changer malfunctions. Notably, these performance challenges persisted even for participants like \deleted{P10}\added{P9}, who possessed a high-performance system equipped with a 12th generation Intel processor, underscoring the severity of the problem.

Besides off-the-shelf voice changer software, \deleted{P12}\added{P11} used an open-source AI voice changer, named W-okada \cite{w-okada_voice_changer}. While having extensive programming background, \deleted{P12}\added{P11} still found the voice changer overly-complex. W-okada required her to follow a YouTube tutorial, since the instructions for setup were in Japanese. Furthermore, she had to collect voice samples for training the AI, deal with command-line interfaces, and train each AI voice, which took over 100 hours. The results of this extensive work, however, were immediately appealing to the researcher---when the participant entered the interview, the researcher was unaware that the participant was using a voice changer. In addition, the relatively easy task of blending multiple AI voices together using the voice changer allowed \deleted{P12}\added{P11} to create a unique and humanlike voice. W-okada, unlike other AI voice changers, runs on the GPU. With an RTX 3060 12GB, the participant was able to both run VRChat and the voice changer at the same time without straining her system's resources.

\paragraph{Paywalls}
Many of the most powerful voice modification tools, such as VoiceMod Pro and MorphVox Pro, are locked behind subscription services and paywalls. For transgender users, many of whom may not be able to afford expensive subscription services, the lack of powerful free options forces them to settle for less equipped voice changers. Five participants \deleted{(P2, P3, P8, P12, P14)}\added{(P1, P2, P7, P11, P13)} mentioned that paywalls or subscription services stopped them from trying fully-featured versions of voice changers. \deleted{P14}\added{P13} highlighted how the companies behind these voice changers understand the market, and lock gender-swapping features behind paywalls:
\added{``}\textit{They didn't offer any good gender changing options in the free version, because of course they didn't. They knew what people wanted to use it for. And they knew how they could get people to pay.}\added{''}
\added{\paragraph{Clocking}
While most participants reached relatively natural-sounding voices with their voice changers, five (P3, P5, P6, P8, P9) referenced ``clockable'' voice changers, that is, voice changers that others could tell were in use, as an issue. P5 and P9 mentioned accidentally using their unmodified voices when the voice changer was not enabled as a factor which could out them to others, with P9 mentioning a clear attitude shift when others heard her unmodified voice. For P3, the imperfect nature of the voice changer pushed her to pursue voice training, since she was jealous of others who did not have to worry about the voice changer giving away their identities:}
    \added{``\textit{I would say that I pass on my style because whenever I use voice changers I would get addressed correctly. Until some people got really smart and started figuring out it was a voice changer. And that's one of the main reasons I wanted to stop too, because I started to get a little bit more jealous of people that were able to mark their voice using these voice mechanics, these voice skills [voice training] that I mentioned before.}''}
\section{Discussion}
We present the first study to examine voice changers as a solution to TGNC users' issues with voice-based dysphoria, misgendering, and harassment in social VR. We identified four reasons why TGNC individuals use voice changers in social VR (Section \ref{sec:findings:motivations}), These reasons include (1) as an alternative to difficult voice training, (2) as a protective measure against harassment, (3) to align their voice with their gender presentation, and (4) to achieve a sense of personal comfort and confidence when speaking with others. We also found that voice changer use affirmed TGNC users' identities by enabling them to pass, and, for some, played a role in their exploration of their gender identities (Section \ref{sec:findings:social_affirmation}). A notable impact of passing as cis was that TGNC users experienced swapped experiences of gender biases and sexism (Section \ref{sec:findings:sexism_preferential_treatment}). \yuhang{ANSWERED: convert all section numbers to ref}

We found that TGNC individuals had mixed views of how voice changers interacted with voice training. Some users felt that voice changers acted as a \deleted{crutch}\added{roadblock}, making voice training more difficult to pursue. \added{Others felt that the voice changers provided necessary support when they couldn't or didn't want to voice train, but could cause overreliance, acting as a ``crutch.''} \yuhang{ANSWERED: modify; distinguish roadblock and crutch.} \deleted{Other}\added{Still, many} users recognized voice changers as a springboard for voice training, providing a supporting role during the voice training process, both as voice ``training wheels'' by allowing individuals to safely and incrementally alter their voice, and as a voice training goal (Section \ref{sec:findings:changers_and_training}).

\added{Besides social impact}, TGNC voice changer users noted that voice changing software had various technical problems and limitations. Settings were vague and confusing, AI voice changers were either too resource intensive---or in the case of W-Okada, impossible to use without programming knowledge---and regularly locked behind paywalls and subscriptions (Section \ref{sec:findings:limitations_of_voice_changers}).

In this section, we discuss the unique possibilities offered by voice changers as a supplement to avatars for gender exploration in social VR, design implications for creating \added{effective and user-friendly} voice changers that meet TGNC users' needs, \added{and considerations for leveraging voice changers to support voice training}. 

\subsection{Gender Exploration via Voice Changers in Social VR}
Previous work evaluated avatars as a tool for gender exploration and affirmation, but noted that the main form of communication, voice chat, posed risks of outing TGNC users to others \cite{rediscovering_the_body_online_trans}. Our findings regarding participants' cis-passing experiences illustrate how the combination of social VR avatars and voice changers can effectively simulate gender presentations that might otherwise require costly and labor-intensive medical transitions and voice training in real life. This lower-effort and accessible solution to changing others' perception of one's gender poses unique benefits for transgender identity exploration. While previous studies have explored the impact of avatar embodiment effects on identity and gender exploration \cite{avatar_embodiment_games, body_avatar_and_me_perception_of_self}, these studies are limited to an individual's perceptions of one's body, and do not explore the impacts of being treated as the avatar's gender in the social VR environment.

Multiple participants said that using a voice changer strengthened their decision to transition, since they enjoyed the social aspects of presenting as the gender they aligned their voice and avatar with. These virtual experiences differ from typical first-experiences of attempting social transition, where hurdles such as difficulty with getting gendered appropriately by others, being treated as something separate from the goal gender (e.g., getting treated as a transgender woman rather than just as a woman), and the stigma of being transgender hinder gender exploration. External factors, such as social stigma and non-acceptance are major drivers in cessation of transition \cite{detransition_factors}. By providing an idea of what an end-game for transitioning could be like, virtual experiences of passing can be profoundly affirming and supportive, providing individuals early in transition with the opportunity to build confidence in their decision to transition. Social VR, when used with a voice changer, offers a unique platform for transgender identity exploration that complements traditional real-world experiences, ultimately contributing to a more informed and empowered decision-making process for individuals on their gender journey.
\subsection{Design Implications for Voice Changer \deleted{Accessibility}\added{Usability}}
Our study identified several key challenges around the use of voice changers, including the scarcity of available voice presets, the intricate nature of voice changer setup, the performance instability arising from system resource constraints, and the affordability of subscription-based models for accessing essential features. These challenges highlight areas where improvements can be made to enhance the \deleted{accessibility}\added{usability} of voice changers for TGNC users.

\paragraph{Designing better foundations} 
Voice changing software presets often revolve around generic character archetypes, featuring options like anime-inspired female personas or stereotypical dorky male voices. Unfortunately, these presets fall short of meeting the unique needs of transgender users who are in search of voice modifications that truly reflect their individual identities. Two of the voice changers used by participants avoided this issue in different ways, but were imperfect in other ways. The neural network voice transformer found in MorphVox allows transgender users to maintain certain elements of their original voices while achieving desired modifications, but the sliders for further customization are complex and difficult to understand. W-Okada, which enables users to blend multiple AI voices to craft a personalized and unique voice, offers transgender users a practical means of establishing realistic voice goals and fostering a stronger sense of identity affirmation, but is extremely difficult to use given \deleted{its}\added{the} command-line interface\added{s required for voice model generation and merging} and lack of documentation. \added{While not a real-time solution, AVOCUS, by Byeon et al., presents an alternative voice alteration system that enables target voice searching with an understandable hashtag system and customization based on the user's own voice, preserving vocal features \cite{avocus}. To support more effective voice modification, developers should consider integrating design foundations produced by prior research (e.g., AVOCUS) alongside the specific needs of target groups, such as TGNC users, to produce more tailored tools.} 

\paragraph{Keep it simple}
Our findings highlight the challenges encountered by numerous transgender users when attempting to create convincing human-like voices using voice changers equipped with an overwhelming array of options. Not to mention that our recruitment criteria limited our ability to capture the experiences of transgender individuals who faced difficulties in developing a functional voice and subsequently abandoned their efforts. Future voice changer design should simplify the voice adjustment process by limiting the number of parameters; abstracting features into understandable categories for transgender users that mimic aspects of voice training, such as voice pitch and resonance/weight; involving neural-network training on users' voices for best-fit approximations of gender swaps, such as in MorphVox; or providing AI presets and allowing users to merge them using a single weight value for each.

\paragraph{Addressing processor constraints}
AI voice changers, despite their capacity to generate highly realistic human voices, can exert significant pressure on the CPU. This strain on the CPU becomes problematic, particularly in social VR applications such as VRChat, where the CPU is responsible for managing numerous users with intricate avatars within the virtual environment. Our research findings indicate that GPU-based AI voice changers, exemplified by W-Okada, do not encounter this problem. Since individuals employing PCs for social VR typically possess gaming-grade GPUs, directing the development of future AI voice changers towards GPU compatibility should effectively address this concern. \added{Still, developers should keep in mind the hardware available to users and the types of software that run alongside their voice changers. When the application is less strenuous, such as voicechat or videochat applications (e.g., Discord, Zoom), CPU voice changers may be the optimal choice, since they do not assume that users have access to powerful GPUs, but in more CPU-intensive applications (e.g., VRChat), developers should consider adding GPU-compatible modes.} 

\paragraph{A community-driven solution}
Developers lock gender-swapping settings in voice changers behind paywalls, and transgender users are regularly not in positions to afford expensive voice changer subscriptions. A community-driven solution, integrating voice model crowdsourcing networks like AI Hub, a Discord community for sharing voice models, alongside a piece of open-source software, like W-Okada, would satisfy the needs of these users, while also providing a platform for researchers to build on.


\subsection{Voice Changers as a Support for Voice Training}


\added{
Voice training is a vital approach for TGNC users to align their voice with their gender identities. Given the arduous process of voice training and difficulties with measuring intermediary success, TGNC individuals have developed communities, applications, and online video-based and text-based guides for voice training techniques. Historically, modern media technologies have been integral to sharing insights on transgender voice training, with ``Melania Speaks!'' providing the first full series on transgender voice training on VHS and DVD in the 1990s \cite{melanie_speaks}. New methods range from voice training video series on YouTube \cite{transvoicelessons_youtube}, Discord communities for voice training such as "TransVoice" and "Scientifically Augmented Voice" \cite{discord_transvoice_2021}, and guides on Reddit for attaining specific goals \cite{reddit_transvoice}, to full-fledged mobile applications which automatically determine feminine, masculine, and androgynous qualities of their voices \cite{voice_training_app, open-source_voice_training_app}. In our study, we identified social VR as a new avenue for voice training, with voice changers as a possible support, crutch, and roadblock.}

\added{
As potential support, participants reported a range of experiences with voice changers in connection to their voice training goals. Some used the modified versions of their own voices from the voice changer as a step towards their vocal targets. Others found that using voice changers motivated them to start formal voice training. However, some chose to stop using voice changers and instead used social VR as a platform for practicing their voice. There were also those who combined the use of voice changers with voice training, combining alterations in their voice changer with alterations in their voice to avoid suspicion from others. Based on these findings, a future work could explore voice changing technology in combination with social VR as an important addition to modern transgender voice therapy.}
\added{
\paragraph{Social VR as a Safe Environment for Passive Training}
While useful for exercising vocal muscles and ascertaining the feeling of a target voice, the exercises supported by modern voice training media, applications, and community events focus primarily on active voice training, where most if not all cognition is directed towards maintaining the desired voice \yuhang{ANSWER: added a reference later and broke it up into multiple sentences to better frame it -- reference}. \added{In contrast,} passive voice training requires using the trained\added{, intermediary} voice in day-to-day communication, a task which normalizes practicing the new voice in parallel with performing other cognitively challenging tasks, which is a major barrier for voice training to many individuals and requires additional techniques, such as ``voice-reseting'' \cite{voice_training_in_conversation, maintaining_deeper_voice_conversation}. Passive voice training is important since it regularizes consistently using a trained voice, and thus can overwrite vocal habits of the previous voice \cite{BeckerENT2024Transfeminine}. However, this would typically force TGNC individuals to come out to others in real life, since real-world interactions would be the primary forum for passively training. Being out as a TGNC person, especially with a non-passing voice, poses enormous risks, including social, economic, and physical risks \cite{glaad_unsafe_america_2023} \yuhang{ANSWERED: broken reference}. While still not free from risks of harassment, social VR environments provide a virtual emulation of real-world interactions in which TGNC individuals can practice passive voice training without the risks of coming out in real life.
\paragraph{Voice Changers as a Safety Net}
Voice changers provide a way to address the concerns of others hearing an intermediary, non-passing voice during voice training in social VR by altering the voice others hear to be more masculine or feminine. This allows TGNC users to avoid harassment in public virtual worlds while also maintaining comfort in that the voice that others hear represents themselves. By separating voice presentation from practice, such as by having others hear the modified voice while the practicing individual hears their trained voice, TGNC individuals engaging in voice training can enjoy the best of both worlds.
\paragraph{An Iterative Design for Transgender Voice Therapy}
Combining the findings from our study alongside modern voice training techniques, we recommend a framework for voice training to be explored in future work, consisting of (1) using voice changers for discovering a participant's target voice, (2) using voice changers to modify and improve the participant's voice to the target voice, (3) comparing the current trained voice with the target voice to develop statistics to find approachable iterations for voice training in areas such as pitch and resonance, and (4) having participants practice their new voice in a social VR environment, all while providing live statistics and feedback and enhancing their voice with a voice changer to avoid harassment and increase participant comfort. This approach would (1) make voice training more approachable by presenting steps the participant could take in achieving their goal voice, instead of presenting the process as a single journey and (2) allow participants to manage their safety and comfort more effectively in passive voice training scenarios, making passive voice training more accessible to participants who do not want to or cannot present their identities in real life.
}
\subsection{Limitations and Future Work}
This study has several limitations that warrant consideration. Our participants exclusively used VRChat, and as such, the generalizability of their experiences to other social VR platforms like RecRoom and Horizon Worlds remains uncertain. Future research should endeavor to explore a more diverse range of social VR platforms to determine if the findings hold true across different virtual environments.
Moreover, the study featured a limited number of participants, which inherently restricts the scope of the study and potentially limits its applicability to a broader user base. To address this limitation, a future research endeavor could extend the study duration and recruitment efforts to include a larger and more diverse sample of users who employ voice changers for gender identity presentation.

Another noteworthy limitation is that our study did not delve into the experiences of TGNC users who abandoned the use of voice changers and the reasons behind their decisions. A subsequent research project could adopt a survey-based approach designed around our findings while widening the eligibility criteria \added{(e.g., including participants who used or tried but abandoned voice changers)}. This approach would provide a more comprehensive understanding of how to design inclusive and \added{easy-to-use} voice changers that cater to the needs of TGNC users effectively.

\added{
An additional limitation of our study is the exclusive use of qualitative methodology. While this approach allowed us to cover nuanced personal and social phenomena as they occured in our exploratory research, the lack of quantitative data prevents us from providing measurable, statistical insights into the use of voice changers by TGNC users in social VR. Future research could benefit from a mixed-methods approach, integrating quantitative data such as online surveys to create a more holistic understanding of the role and impact of voice changers for TGNC users in social VR.
}

Moreover, participants in our study recounted past experiences, which might have resulted in some details and difficulties related to voice changer setup being overlooked or inaccurately remembered. Future research could involve participants who are new to voice-changing technology and assign them the task of setting up a voice changer to gain more precise insights into the setup process and associated challenges.

Finally, participants primarily remembered salient and recognizable behaviors of others when recalling experiences of harassment, misgendering, or other negative encounters tied to using their unmodified voices in social VR. Future research efforts could adopt an alternative approach by capturing these experiences through participants recording their social VR sessions while using and not using voice changers, similar to those employed in Zhang et al. \cite{impact_of_disability_signifiers}. This method would provide a more comprehensive and detailed perspective on the challenges faced by users in real-time scenarios. \yuhang{in the IEEE VR context, this study may not be that interesting... replace it with limitations in the lack of quantitative research, so that the limitation section won't be too long.}
\section{Conclusion}
In this paper, we conducted interviews to delve into the experiences of TGNC users employing voice changers in social virtual reality (VR) environments. Our findings have shed light on several critical aspects, including the diverse goals that TGNC individuals aim to achieve with voice changers, the marked disparities in their experiences with and without these voice-modifying tools, the profound impact of ``passing'' as cisgender in social VR for those who may not have this opportunity in real-life settings, attitudes around the complex relationship between voice training and voice changers, and the \deleted{prevalent challenges faced during voice changer setupand performance}\added{limitations of current voice changers for TGNC users}.

\deleted{Our findings contribute to the existing literature on how social VR avatars and voice changers empower TGNC users to authentically reflect their gender identities, offering invaluable opportunities for gender exploration and affirmation. Furthermore, our insights into the challenges related to voice changer interfaces can inform the design of future voice changer technologies to be more effective and user-friendly.}

\acknowledgments{
\added{This work was supported in part by the National Science
Foundation under Grant No. IIS-2328182.} }

\bibliographystyle{abbrv-doi}

\bibliography{template}

\appendix
\renewcommand\thefigure{\thesection.\arabic{figure}}    
\setcounter{figure}{0}  
\renewcommand\thetable{\thesection.\arabic{table}}    
\setcounter{table}{0}  
\section{Appendix}
\label{sec:appendix}
\subsection{Pre-Screening Questionnaire}
\label{appendix: screen}
Respondents were eligible to participate in our study if they were 18 years old or above, had frequent experience with social VR applications such as VRChat,
identified as transgender and/or gender non-conforming, and had experience with using voice-changing technology in social VR.
\begin{enumerate}
    \item How old are you?
    \item Do you identify as transgender and/or gender non-conforming?
    \item How often do you use social VR applications such as VRChat?
    \item Do you use voice-changing technology in social VR, such as voice modifiers, synthesizers, or changers?
\end{enumerate}

\subsection{Interview Questions}
\label{sec: interview}

\subsubsection{Demographic Questions}
\begin{itemize}

\item What is your age?
\item What is your ethnicity or racial background?
\item Do you identify as transgender?
\item Do you identify as gender non-conforming?
\item Are there any other aspects of your identity that you would like to mention?
\item Are you satisfied with your real-life voice? Why?
If not:
\begin{enumerate}
    \item Are you doing anything to change your voice?
\end{enumerate}
\item What voice modifier do you use?
\item When do you use a voice modifier?
\item When you use your voice in public, do others gender you as you wish?
\item What is your ideal voice?
\item Have you encountered any barriers in real-world social activities because of your gender identity?
If yes:
\begin{enumerate}
    \item Are there any voice-related barriers?
    \item Can you give me some examples of memorable experiences?
    \item Do you have strategies to overcome these barriers?
\end{enumerate}

\end{itemize}


\subsubsection{Virtual World Experiences}

\begin{itemize}
\item What social VR platforms do you typically use?
\item Which do you use most frequently?
\item How often do you use social VR?
\item What do you usually do in social VR?
\item Who do you hang out with in social VR?
\item Do you typically interact more with friends on social VR platforms, or do you actively seek out new connections with other users?
\item What do you enjoy about social VR?
\item Can you show me your avatar or set of avatars that you use?
\item Are there any specific aspects of your avatar that you would like to highlight?
\item Do you use your avatar to express your gender identity?
If yes:
\begin{enumerate}
    \item How?
    \item How effective is your avatar in expressing your gender identity?
\end{enumerate}
If no:
\begin{enumerate}
    \item Why?
\end{enumerate}
\item Do people treat you differently depending on the avatar you use?
If yes:
\begin{enumerate}
    \item Why do you think this is?
\end{enumerate}
\item Are there any benefits to using social VR as a transgender/gender non-conforming user?
\item Are there any barriers you’ve faced in social VR as a transgender/gender non-conforming user?

\end{itemize}

\subsubsection{Voice and Voice Modification Experiences in Social VR}

\begin{itemize} 

\item When not using a voice modifier, do you use voice chat in social VR?
If yes:
\begin{enumerate}
    \item How often do you use it?
    \item Are there any other ways that you communicate with others?
\end{enumerate}
If no:
\begin{enumerate}
    \item Why?
    \item Do you use alternative methods to communicate with others?
\end{enumerate}
\item What situations do you typically use your unmodified voice in?
\item Why?
\item Who do you typically use your unmodified voice with?
\item Why?
\item Do you encounter any barriers when you use your real voice?
If yes:
\begin{enumerate}
    \item What are your most memorable stories?
    \item How did you feel?
\end{enumerate}
\item What strategies do you use to avoid problems with using your voice?
\item What are the most effective solutions?
\item Why?
\item What are the drawbacks of the solutions?
\item When it comes to using voice modifiers in social VR, what initially sparked your interest or curiosity in experimenting with them?
\item What voice modifiers do you use?
\item Are there any features that you enjoy from these voice modifiers?
\item How frequently do you use a voice modifier?
\item In what situations do you use a voice modifier?
\item Does voice modifier use vary between groups – such as friends vs. strangers? How? Why?
\item Does your voice modifier use vary by platform – either by which social vr environment you’re in or the specific hardware you’re using? How? Why?
\item Does your voice modifier use vary by avatars? How? Why?
\item What voice do you generate with your voice modifier?
\item Can you show me?
\item Do you use a single voice or multiple voices?
\item Why?
\item Do you use any AI or preset voices?
\item Which ones?
\item Why?
\item Do you like or dislike such a feature?
\item How would you improve it?
\item Do you customize the voice modifier directly?
If yes:
\begin{enumerate}
    \item How did you tune your voice modifier?
    \item Why did you tune it in that way?
    \item How easy or hard was it to tune your voice modifier?
    \item What were the challenges you faced?
    \item How do you want to improve it?
\end{enumerate}
\item When using social VR, do you go back to the voice modifier interface to change your voice?
If yes:
\begin{enumerate}
    \item How does this impact your experience with social VR?
\end{enumerate}
\item Why do you use a voice modifier in social VR?
If gender presentation is mentioned:
\begin{enumerate}
    \item How effective is a voice modifier for conveying your gender identity?
\end{enumerate}
\item Are there any social barriers that using a voice modifier has created?
\item Have you used a voice modifier to experiment with other gender presentations, such as when figuring out your gender identity?
\item Do you find that others treat you differently when you use a voice modifier to change the perceived gender of your voice?
If yes:
\begin{enumerate}
    \item In what situations?
    \item What differences have you noticed in terms of engagement, responses, or overall social dynamics?
\end{enumerate}
\item How does using a voice modifier change how you feel about yourself?
Prompts:
\begin{enumerate}
    \item Self-confidence
    \item Self-esteem
\end{enumerate}
\item How does using a voice modifier impact your behaviors when interacting with others?
Prompt:
\begin{enumerate}
    \item Does it impact your social strategies -- What you feel comfortable and capable of doing?
\end{enumerate}
\end{itemize}


\subsubsection{Technical Questions and Brainstorming Session}

\begin{itemize}

\item Have you ever faced any technical difficulties or limitations with voice modification technology in social VR?
\item Can you describe these difficulties and how they impacted your experience in social VR?
Prompts:
\begin{enumerate}
    \item Finding the right voice modifier
    \item Voice modifier installation
    \item Platform compatibility
    \item Parameter tuning
    \item Real-time usage
\end{enumerate}
\item Are there any improvements or changes you would like to see in voice modification technology for social VR based on your experiences?
\item Do you use a voice modifier on any non-social-vr platforms?
If yes:
\begin{enumerate}
    \item How does your use of voice modification differ compared to social vr on these platforms?
\end{enumerate}
\item Imagine that your favorite voice modifier works with VRChat to design a feature that allows you to change your voice modifier options 
on the fly within the application. What sort of method do you feel would work best for this customization interface? Why?
Prompts:
\begin{enumerate}
    \item Avatar-based (voice is connected to avatar)
    \item Custom sliders (user can modify values like pitch, reverb, etc. directly)
    \item Choosing from a panel of voices (user can select from voice presets, such as AI voices)
\end{enumerate}
\end{itemize}
\added
{
\subsection{Codebook Themes and Subthemes}}
\begin{table*}[H]
    \footnotesize
    \centering
    \caption{\added{Full codebook used for data analysis, including themes, subthemes, and relevant codes}}
    \label{tab:theme}
    \begin{center}
    \begin{tabular}{p{3.5cm}p{3.5cm}p{10cm}}
    \toprule
    \textbf{Theme} & \textbf{Subtheme} & \textbf{Codes}\\
    \midrule
    Voice Changer Use Demographics & Type of Voice Changer Used & Voicemod, Voicemod Persona, Voicemod Pro, w-okada, Clownfish, MorphVox Junior, MorphVox Pro, SuperVoiceChanger \\
\hline
 & Frequency of and Factors Influencing Voice changer Use & selective voice changer usage, voice changer all the time, voice changer use outside of social VR, voice changer use: based on mood, voice changer use: trans people, talking to people with voice changer, using voice changer with streaming, voice changer: even with close friends, voice changer: for fun with close friends, voice changer: for role-play, voice changer: for vocal dysphoria, voice changer: independence from avatar, voice changer: little change with friends, voice changer: lowering frequency of use, voice changer: use as vtuber, voice changer: using for fun, voice changer: when not with close friends \\
\hline
 & Voice Changers Usage and the Social Environment & no social barriers with friends, only use unmodified voice with closest friends, personal comfort in who to use voice changer with, social VR: avoiding strangers, social VR: meeting new people to play games, unmodified voice use: trust, unmodified voice use: small groups, unmodified voice use: with trans users, unmodified voice: only with close friends, voice changer: when not with close friends \\
\noalign{\hrule height 2pt}
Motivations for using modified voice & Less effort for gender presentation compared to voice training & social VR benefits: less effort to express self in social VR, social VR benefits: less effort to pass in social VR, voice training problems: self-confidence, voice training problems: physical effort, voice changer: less effort to present identity \\
\hline
 & Unmodified Voice as source of harassment & harassment over unmodified voice, identity and voice do not mesh, unmodified voice: lack of confidence in voice, mental barrier without voice changer, mental dissociation when mute, priority of voice chat, stigma without voice changer, unmodified voice: avatar-voice gender mismatch, unmodified voice in social VR: avoiding talking, unmodified voice use: difficult to exist in social VR, unmodified voice: fear of others hearing voice, unmodified voice: avoiding others, unmodified voice: used rarely \\
\hline
 & Using voice changer to avoid harassment & using voice changer to avoid stigma, voice changer avoids misgendering, voice changer: acceptance from others, voice changer: difference in treatment, voice changer: improved social experience, voice changer: social acceptance \\
\hline
 & Avatar-Voice Matching & pairing voice changer and avatar, voice changer + avatar: expressing identity, voice changer + avatar: helps with anxiety, voice changer: enhanced via avatar design \\
\hline
 & Confidence and Comfort & comfort with voice, voice changer + avatar: helps with anxiety, voice changer: increasing confidence, voice changer: life-saving, voice changer: alleviates dysphoria \\
\noalign{\hrule height 2pt}
Social Affirmation and Passing & Being percieved as cisgender & voice changer: presenting as cis, voice changer: easier route to affirmation, different kind of harassment when passing \\
\hline
 & Bolstering Confidence in Identity & used voice changer before figuring out trans, voice changer playing a minor role in gender exploration, voice changer: used for fun before transition, voice changer: used for role-play before transition \\
\noalign{\hrule height 2pt}
Sexism and Preferential Treatment & Experiences of Sexism when using voice changer & voice changer: treated in sexist way presenting as female, voice changer: privy to sexism when presenting as male \\
\hline
 & Experiences of Preferential Treatment when using voice changer & voice changer: preferential treatment when presenting as female, voice changer: being one of the guys, voice changer: treated as one of the girls \\
\hline
 & Experiences of Sexism when not using voice changer & unmodified voice: sexism when voice is read as female \\
\hline
 & Experiences of Preferential Treatment when not using voice changer & unmodified voice: preferential treatment when voice is read as female \\
\noalign{\hrule height 2pt}
Interactions between Voice Changers and Voice Training & Voice Changer becomes own voice & having own voice, voice changer: representing self with voice, voice changer: crafting a realistic voice, voice changer: settling into a voice, voice changer: internalization of modified voice, voice changer: replaces unmodified voice as accepted original, voice changer: causing changes to speech patterns, voice changer: changes self-perception, voice changer represents ideal voice \\
\hline
 & Voice Changer as support for training & voice changer and voice training together, voice changer: does not impact active voice training, voice changer: push towards voice training \\
\hline
 & Voice Changer as roadblock for training & social VR as voice-practice space, voice changer is barrier to training voice, voice changer: hinders passive voice training \\
\hline
 & Voice changer as a Crutch for training & voice changer as crutch, using voice changer when lazy, using voice changer when sick, voice changer as fallback to voice training, voice changer as backup to voice training \\
\bottomrule
\end{tabular}
    \end{center}
\end{table*}

\begin{table*}[H]
    \footnotesize
    \centering
    \caption{\added{Continued: Full codebook used for data analysis, including themes, subthemes, and relevant codes}}
    \label{tab:theme2}
    \begin{center}
    \begin{tabular}{p{3.5cm}p{3.5cm}p{10cm}}
    \toprule
    \textbf{Theme} & \textbf{Subtheme} & \textbf{Codes}\\
    \midrule
Voice changer Limitations & Difficulties with Settings and Presets & people finding w-okada too technical, previous voice changers didn't work, w-okada: time-consuming and technical voice changer setup, voice changer: issues with updates -- losing settings, voice changer: technical issues, long setup time for w-okada, cannot describe how voice changer works, changing when sounding robotic, checking that the voice changer is working well, complex voice changer setup, voice changer + vrchat performance issues, system strain coping strategies, following instructions to setup voice changer, frustration with setting up voice changer, high-performance system, itterating on voice changer, not changing voice in middle of social VR, pc-compatible voice changer, people finding w-okada too technical, voice changer performance issues, speech-text-speech difficulty, voicemod persona: voice latency, w-okada instructions in Japanese \\
\hline
 & Performance issues with AI Voice Changers & voice changer performance issues, voice changer + vrchat performance issues, text-to-speech too robotic, voicemod: high computational power, voicemod persona: computational burden, voice changer issues: too robotic/tinny, voice changer keeping up, voice changer: performance issue avoidance strategies, voicechanger: risk of unmodified voice being heard as result of system strain \\
\hline
 & Paywalls & didn't try paid-for voice changers, limitations of free-version voice changers, voicemod: switching voice on you in free version, voicemod free version problems, voice changer wishes: free \\
\hline
 & Clocking & voicemod persona: easy to clockvoice changer: embarassment of accidentally having it off, voice changer: indicates trans identity, voice changer: attitude change after hearing real voice, voice changer: clocking \\
\bottomrule
    \end{tabular}
    \end{center}

\end{table*}

\clearpage

\added{
\subsection{Various Voice Changer Interfaces}}
\begin{figure*}[!ht]
    \centering
    \begin{subfigure}{0.6\textwidth}
        \includegraphics[width=\textwidth]{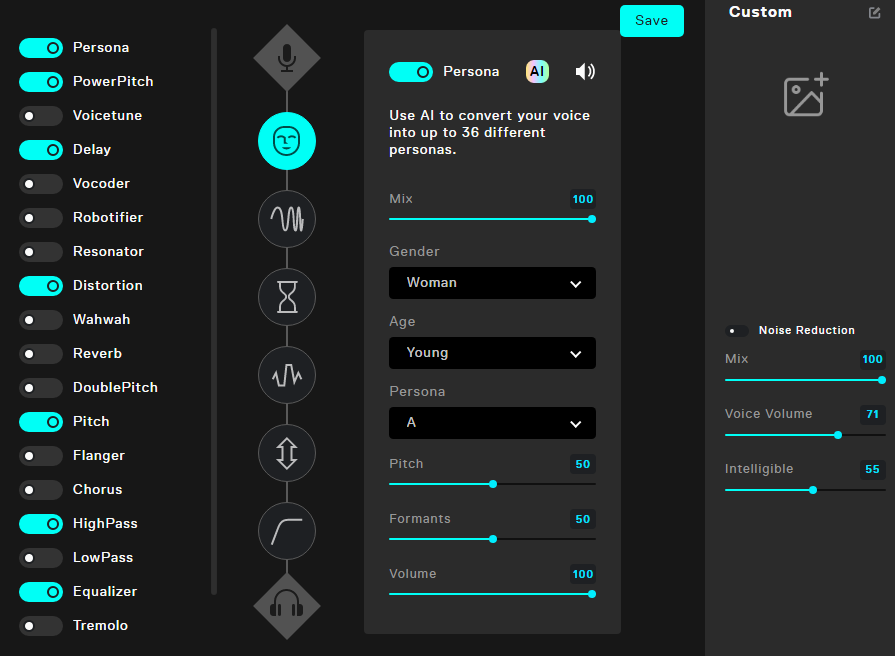}
        \caption{Voicemod}
        \label{fig:voicemod}
    \end{subfigure}

    \begin{subfigure}{0.6\textwidth}
        \includegraphics[width=\textwidth]{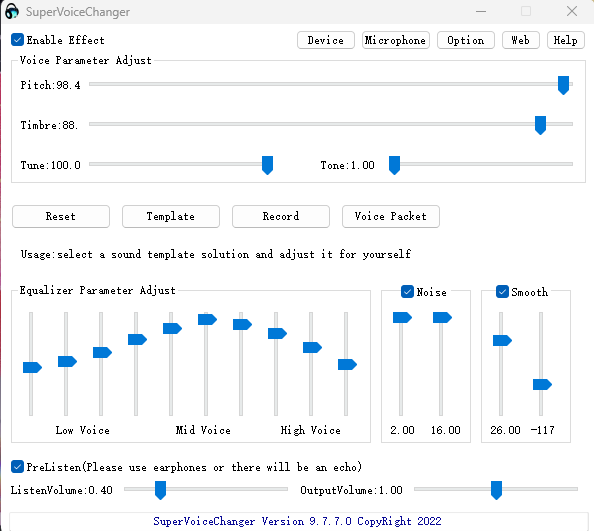}
        \caption{SuperVoiceChanger}
        \label{fig:supervoicechanger}
    \end{subfigure}
    \caption{\added{Interfaces of various voice changer software: Voicemod Voicelab and SuperVoiceChanger}}
\end{figure*}

\begin{figure*}[!ht]
    \vspace{5mm}
    \begin{subfigure}[b]{0.45\textwidth}
        \includegraphics[width=\textwidth]{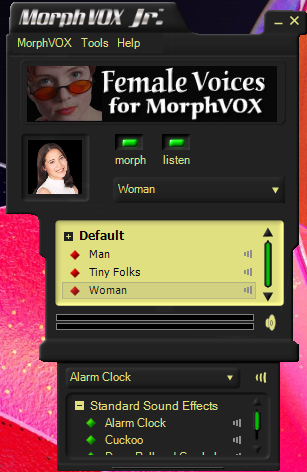}
        \caption{MorphVOX Junior}
        \label{fig:morphvox}
    \end{subfigure}
    \hfill
    \begin{subfigure}[b]{0.45\textwidth}
        \includegraphics[width=\textwidth]{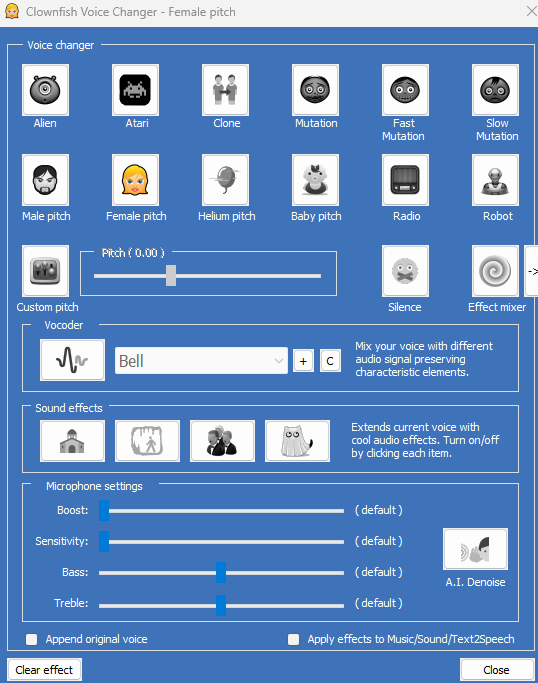}
        \caption{Clownfish}
        \label{fig:clownfish}
    \end{subfigure}

    \vspace{5mm}
    \begin{subfigure}[b]{0.4\textwidth}
        \includegraphics[width=\textwidth]{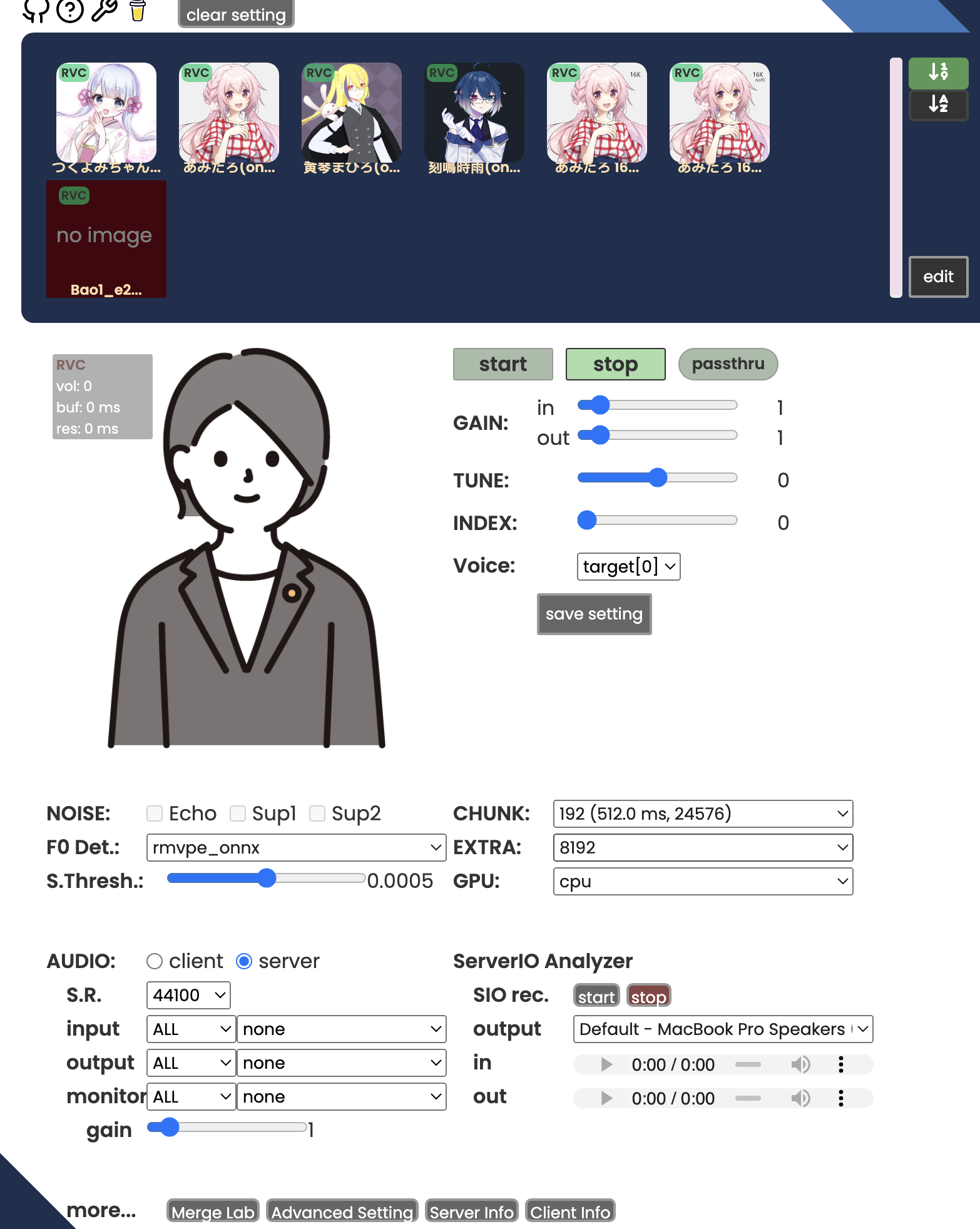}
        \caption{W-Okada}
        \label{fig:wokada}
    \end{subfigure}
    
    \caption{\added{(Continued) Interfaces of various voice changer software: MorphVOX Junior, Clownfish, and W-Okada.}}
    \label{fig:voicechangers}
\end{figure*}

\end{document}